\documentclass[twocolumn,showpacs,preprintnumbers,amsmath,amssymb]{revtex4}
\usepackage{graphicx}
\usepackage{dcolumn}
\usepackage{bm}
\begin{document}
\newcommand{\hs}{\hspace*{0.5cm}}
\newcommand{\vs}{\vspace*{0.5cm}}
\newcommand{\be}{\begin{equation}}
\newcommand{\ee}{\end{equation}}
\newcommand{\bea}{\begin{eqnarray}}
\newcommand{\eea}{\end{eqnarray}}
\newcommand{\ben}{\begin{enumerate}}
\newcommand{\een}{\end{enumerate}}
\newcommand{\nn}{\nonumber}
\newcommand{\crn}{\nonumber \\}
\newcommand{\non}{\nonumber}
\newcommand{\noi}{\noindent}
\newcommand{\al}{\alpha}
\newcommand{\la}{\lambda}
\newcommand{\bet}{\beta}
\newcommand{\ga}{\gamma}
\newcommand{\va}{\varphi}
\newcommand{\om}{\omega}
\newcommand{\pa}{\partial}
\newcommand{\fr}{\frac}
\newcommand{\bc}{\begin{center}}
\newcommand{\ec}{\end{center}}
\newcommand{\Ga}{\Gamma}
\newcommand{\de}{\delta}
\newcommand{\De}{\Delta}
\newcommand{\ep}{\epsilon}
\newcommand{\varep}{\varepsilon}
\newcommand{\ka}{\kappa}
\newcommand{\La}{\Lambda}
\newcommand{\si}{\sigma}
\newcommand{\Si}{\Sigma}
\newcommand{\ta}{\tau}
\newcommand{\up}{\upsilon}
\newcommand{\Up}{\Upsilon}
\newcommand{\ze}{\zeta}
\newcommand{\ps}{\psi}
\newcommand{\Ps}{\Psi}
\newcommand{\ph}{\phi}
\newcommand{\vph}{\varphi}
\newcommand{\Ph}{\Phi}
\newcommand{\Om}{\Omega}
\def\lappeq{\mathrel{\rlap{\raise.5ex\hbox{$<$}}
{\lower.5ex\hbox{$\sim$}}}}

\title{ Standard-model-like Higgs boson production at the CERN  LHC in \\
 3-3-1  model  with right-handed neutrinos}

\author{Le Duc Ninh}%
 \email{ ldninh@grad.iop.vast.ac.vn}
 \affiliation{%
Department of Physics, Hanoi University of Education,  Hanoi,
Vietnam }
\author{Hoang Ngoc Long}%
 \email{hnlong@iop.vast.ac.vn}
\affiliation{%
Institute of Physics, VAST, P. O. Box 429, Bo Ho, Hanoi 10000,
Vietnam }

\date{\today}

\begin{abstract}
The models based on $\mbox{SU}(3)_C\otimes \mbox{SU}(3)_L \otimes
\mbox{U}(1)_X$ gauge group (3-3-1) contain new Higgs bosons and
one of them is the SM-like Higgs boson $h$. Production of this
Higgs boson at $p p$ colliders in the framework of 3-3-1 model
with right-handed neutrinos is calculated. We found that the
contribution of the real $Z^\prime$ to the process $pp\to hZ$ is
nearly 60 fb if $M_{Z^\prime}$ is about 1 TeV. The decay width and
the branching ratios of the $Z'$ extra neutral gauge boson are
systematically discussed.
\end{abstract}


\maketitle

\section{Introduction}

The recent experimental results of SuperKamiokande
Collaboration~\cite{superK}, KamLAND~\cite{kam} and SNO~\cite{sno}
confirm that neutrinos have {\it tiny} masses and oscillates. This
implies that the standard model (SM) must be extended.

Among the possible extensions of SM, a curious choice is the 3-3-1
models which are based on the simplest non-Abelian extension of
the SM group, namely, the $\mbox{SU}(3)_C\otimes \mbox{SU}(3)_L
\otimes \mbox{U}(1)_X$~\cite{ppf,flt}. The reason why these models
are appealing have been exposed in many recent
publications~\cite{recent}. The model requires that the number of
fermion families be a multiple of the quark color in order to
cancel anomalies, which suggests an interesting connection between
the number of flavors and the strong color group.
 If further one uses the
condition of QCD asymptotic freedom, which is valid only if the
number of families of quarks is to be less than five, it follows
that $N$ is equal to 3. In addition, the third quark generation
has to be different from the first two, so this leads to the
possible explanation of why top quark is uncharacteristically
heavy.

There are two main versions of the 3-3-1 models as far as lepton
sector is concern. In the minimal version, the charge conjugation
of the right-handed charged lepton for each generation is combined
with the usual $SU(2)_L$ doublet left-handed leptons components to
form an $SU(3)$ triplet $(\nu, l, l^c)_L$.  No extra leptons are
needed and there we shall call such models minimal 3-3-1 models.
There is no right-handed (RH) neutrino in its minimal version.
Another version adds a left-hand anti-neutrino to each usual
$SU(2)_L$ doublet left-handed lepton to form a triplet. This model
is called the 3-3-1 model with RH  neutrinos.

It is well known that the electroweak symmetry breaking in the SM
is achieved via the Higgs mechanism. In the Glashow-Weinberg-Salam
model there are a single complex Higgs doublet, where the Higgs
boson $h$ is the physical neutral Higgs scalar which is the only
remaining part of this doublet after spontaneous symmetry
breaking. In the extended  models there are additional charged and
neutral scalar Higgs particles. The search for the Higgs boson is
an important part of the Fermilab Tevatron experiments  and will
be one of the main fields of study at the CERN LHC collider. The
experimental detection of the $h$ will be great triumph of the SM
of electroweak interactions and will mark new stage in high energy
physics.

Production of the Higgs boson in the minimal 3-3-1 model at the
$e^- e^+$ Next Linear Collider and in CERN Linear Collider (CLIC)
has been considered in \cite{cmt1}. The aim in this paper is to
consider production of the Higgs boson of the SM  at hadron
colliders.

The rest of the paper is organized as follows: In Sec. 2 a brief
review of the 3-3-1  model with RH neutrinos is presented. Sec. 3
is devoted to the decay of the $Z'$ extra gauge bosons. In Sec. 4
we discuss single production of the $Z'$ and the SM Higgs bosons.
Finally, our conclusions are presented in Sec. 4.

\section{A review of 3-3-1  model  with  RH neutrinos}

The 3-3-1 model is based on the gauge group \bea
\mbox{SU}(3)_C\otimes \mbox{SU}(3)_L \otimes \mbox{U}(1)_X.\eea In
the version considered in this paper, one adds a left-hand
anti-neutrino to each usual $SU(2)_L$ doublet left-handed lepton
to form a triplet~\cite{flt}
\begin{equation}
f^{a}_L = \left( \begin{array}{c}
               \nu_L^a\\ e_L^a\\ (\nu_L^C)^{a}
\end{array}  \right) \sim (1, 3, -1/3), e^a_R\sim (1,
1, -1), \label{l2}
\end{equation}
 where $a=1,2,3$ is the generation index.
Two first quark generations belong to  antitriplets and the third
is in triplet
 \be Q_{iL} = \left( \begin{array}{c}
                d_{iL}\\-u_{iL}\\ D_{iL}\\
                \end{array}  \right) \sim (3, 3^*, 0),
\label{q12} \ee \bea u_{iR}&\sim &(3, 1, 2/3), d_{iR}\sim (3, 1,
-1/3),\crn
 D_{iR}&\sim &(3, 1, -1/3),\ i=1,2,\nn
 \eea \be
 Q_{3L} = \left( \begin{array}{c}
                 u_{3L}\\ d_{3L}\\ T_{L}
                \end{array}  \right) \sim (3, 3, 1/3),
\label{q13} \ee
\[ u_{3R}\sim (3, 1, 2/3), d_{3R}\sim (3, 1, -1/3), T_{R}
\sim (3, 1, 2/3).\] The electric charge operator is defined as
\bea Q=\fr{1}{2}\lambda_3-\fr{1}{2\sqrt{3}}\lambda_8 + X.
\label{dt} \eea From {\ref{dt}) we get the electric charge of the
new exotic quarks to be $-\fr 1 3$ and $\fr 2 3$ as in (\ref{q12})
and (\ref{q13}).

\hs The gauge symmetry breaking and fermion mass generation can be
achieved with just three $\mbox{SU}(3)_L$ triplets
\begin{equation}
\eta = \left( \begin{array}{c}
               \eta^{0}\\ \eta^{-}\\ {\eta^\prime}^{0}
\end{array}  \right) \sim (1, 3, -1/3),
\end{equation}
\begin{equation}
\rho = \left( \begin{array}{c}
               \rho^{+}\\ \rho^{0}\\ {\rho^\prime}^{+}
\end{array}  \right) \sim (1, 3, 2/3),
\end{equation}
\begin{equation}
\chi = \left( \begin{array}{c}
               \chi^{0}\\ \chi^{-}\\ {\chi^\prime}^0
\end{array}  \right) \sim (1, 3, -1/3),
\end{equation}
with the following VEVs
\begin{equation}
<\eta>^T=(v,0,0),<\rho>^T=(0,u,0),<\chi>^T=(0,0,w).\label{tbck}
\end{equation}
 Note that the scalars $\eta$ and $\chi$ have the same quantum
numbers.

To be consistent with low energy phenomenology, we have to impose
the following condition \bea w \gg v ,u \label{wvu}. \eea

 The neutral gauge boson $Z$ and the new neutral $Z^\prime$
interact with fermions as follows
\begin{eqnarray}
{\cal L}^{NC}&=&\frac{g}{c_W}\left\{\bar{f}\gamma^{\mu}
[a_{1L}(f)\fr{1-\gamma_5}{2}+a_{1R}(f)\fr{1+
\gamma_5}{2}\right] f Z_{\mu}\nonumber\\
             &+& \left.\bar{f}\gamma^{\mu}
[a_{2L}(f)\fr{1-\gamma_5}{2}+a_{2R}(f)\fr{1+\gamma_5}{2}]f
Z^\prime_{\mu}\right\}\crn
&=&\frac{g}{c_W}\left\{\bar{f}\gamma^{\mu}
[g_{1V}(f)+g_{1A}(f)\gamma_5\right] f Z_{\mu}\nonumber\\
             &+& \left.\bar{f}\gamma^{\mu}
[g_{2V}(f)+g_{2A}(f)\gamma_5]f Z^\prime_{\mu}\right\}.
\label{ncva}
\end{eqnarray}
where $g_{1V}(f)=(a_{1L}+a_{1R})/2, g_{1A}(f)=(a_{1R}-a_{1L})/2,
g_{2V}(f)=(a_{2L}+a_{2R})/2$ and $g_{2A}(f)=(a_{2R}-a_{2L})/2$ are
defined as \cite{hnl} \bea
a_{1L,R}(f)&=&T^3(f_{L,R})-s^2_WQ(f),\crn
a_{2L,R}(f)&=&-c^2_W\left[\fr{3X(f_{L,R})}{\sqrt{3-4s^2_W}}-
\fr{\sqrt{3-4s^2_W}}{2c^2_W}Y(f_{L,R})\right],\crn\eea where
$Y=2X-\lambda_8/\sqrt{3}$. The couplings are presented in Table
\ref{zff} and Table \ref{zpff}.
\begin{table}[h]
\caption{
     The couplings of $Z$ with fermions}
\begin{center}
\begin{tabular}{|c|c|c|}  \hline
f &$ g_{1V}(f)$ & $g_{1A}(f)$  \\  \hline $e, \mu, \tau$  &
$-\frac{1}{4}+s_W^2$&$\frac{1}{4}$\\  \hline $\nu_{e}, \nu_{\mu},
\nu_{\tau}$ &$\frac{1}{4}$ & $-\frac{1}{4}$ \\ \hline u,c,t&$
\frac{1}{4}-\frac{2s^2_W}{3}$&$-\frac{1}{4}$\\ \hline
d,s,b&$-\frac{1}{4}+\frac{s^2_W}{3} $&$\frac{1}{4}$\\ \hline
T&$-\frac{2}{3}s^2_W$&$ 0$\\
\hline
$D_i$&$\frac{1}{3}s^2_W$&$0$\\
\hline
\end{tabular}
\label{zff}
\end{center}
\end{table}
\vs

\begin{table}[h]
\caption{
     The couplings of $Z'$ with fermions}
\begin{center}
\begin{tabular}{|c|c|c|}  \hline
f &$ g_{2V}(f)$ & $g_{2A}(f)$  \\  \hline $e, \mu, \tau$  &
$(-\frac{1}{4}+s_W^2)\frac{1} {(3-4s^2_W)^{1/2}}$&$\frac{1}
{4(3-4s^2_W)^{1/2}}$\\  \hline $\nu_{e}, \nu_{\mu}, \nu_{\tau}$
&$-\frac{(3-4s^2_W)^{1/2}}{4}$ & $\frac{(3-4s^2_W)^{1/2}}{4}$ \\
\hline t &$-\frac{3+2s_W^2}{12(3-4s_W^2)^{1/2}}$&
$(\frac{1}{2}-s_W^2)\frac{1}{2(3-4s_W^2)^{1/2}}$\\ \hline b &
$-\frac{(3-4s_W^2)^{1/2}} {12}$ & $\frac{1} {4(3-4s^2_W)^{1/2}}$\\
\hline u,c&$ (\frac{1}{4}-\frac{2s^2_W}{3})\frac{1}{(3-
4s^2_W)^{1/2}}$&$-\frac{1}{4(3- 4s^2_W)^{1/2}}$\\ \hline
d,s&$(\frac{1}{4}-\frac{s^2_W}{6})
\frac{1}{(3-4s^2_W)^{1/2}}$&$-(\frac{1}{2}-s^2_W)
\frac{1}{2(3-4s^2_W)^{1/2}}$\\ \hline T&$(3-7s^2_W)\frac{1}{6(3-
4s^2_W)^{1/2}}$&$ -\frac{c^2_W}{2(3- 4s^2_W)^{1/2}}$\\
\hline $D_i$&$-\frac{3-5s^2_W}{6(3-
4s^2_W)^{1/2}}$&$ \frac{c^2_W}{2(3- 4s^2_W)^{1/2}}$\\
\hline
\end{tabular}
\label{zpff}
\end{center}
\end{table}
To get interactions among Higgs bosons with the $Z$ and the extra
$Z'$, we consider \bea {\cal
L}_{Higgs}&=&(D_{\mu}\eta)^+(D^{\mu}\eta)+(D_{\mu}\rho)^+(D^{\mu}\rho)
+(D_{\mu}\chi)^+(D^{\mu}\chi)\crn\label{dnh} \eea where the
covariant derivatives of triplets are given by \bea D_\mu\phi
&=&\left(
\partial_\mu+ig\sum_{a=1}^8W_\mu^a\fr{
\lambda_a}{2}+ig_XXB_\mu\fr{\lambda_9}{2}\right)\phi\\
&=&\partial_\mu\phi+ \mbox{diag}(a^1,a^2,a^3)_\mu\phi +M_\mu\phi
\eea where $\lambda_a$, $a=1,\cdots ,8$ are Gell-Mann matrices,
$\lambda_9=\sqrt{2/3}\ \mbox{diag}(1,1,1)$, \bea
a^1_\mu&=&i\fr{g}{2}\left(W^3_\mu+\fr{1}{\sqrt{3}}W^8_\mu
+\fr{g_X}{g}\sqrt{\fr{2}{3}}XB_\mu\right),\crn
a^2_\mu&=&i\fr{g}{2}\left(-W^3_\mu+\fr{1}{\sqrt{3}}W^8_\mu
+\fr{g_X}{g}\sqrt{\fr{2}{3}}XB_\mu\right),\crn
a^3_\mu&=&i\fr{g}{2}\left(-\fr{2}{\sqrt{3}}W^8_\mu+
\fr{g_X}{g}\sqrt{\fr{2}{3}}XB_\mu\right),\label{a1}\\
M_\mu&=&i\fr{g}{\sqrt{2}}\left(\begin{array}{ccc}0 & W_\mu^+&X_\mu^0\\
W_\mu^-&0 &Y_\mu^-\\X_\mu^{0*}&Y_\mu^+&0\end{array}\right),\eea
where the non-self-conjugated gauge bosons $W^+$, $Y^-$, $X$ are
defined as \bea \sqrt{2}\ W^+_\mu &=& W^1_\mu - iW^2_\mu ,
\sqrt{2}\ Y^-_\mu = W^6_\mu - iW^7_\mu ,\nn\\
\sqrt{2}\ X_\mu &=& W^4_\mu - iW^5_\mu. \eea
 The matching
condition gives a relation between coupling constants of two
groups (for the 3-3-1 model with arbitrary beta, see
\cite{pvdhnl}) \bea
\fr{g_X}{g}&=&\fr{3\sqrt{2}t_W}{\sqrt{3-t^2_W}}, \eea where
$t_W\equiv s_W/c_W$, with $s_W\equiv\sin\theta_W$,
$c_W\equiv\cos\theta_W$.

The  gauge bosons are expressed in terms of  the physical ones
through the relation dependent only on  the electric charge form
(\ref{dt}) but not on the Higgs structure \cite{pvdhnl}: \bea
W^3_\mu &=&c_WZ_\mu + s_WA_\mu ,\crn W^8_\mu
&=&\sqrt{1-\fr{t^2_W}{3}}Z^\prime_\mu-\fr{t_W}{\sqrt{3}}\left(c_WA_\mu
- s_WZ_\mu\right),\crn B_\mu
&=&\fr{t_W}{\sqrt{3}}Z^\prime_\mu+\sqrt{1-\fr{t^2_W}{3}}\left(c_WA_\mu
- s_WZ_\mu\right). \eea  Hence we have\bea a^1_\mu
&=&b_1Z^\prime_\mu+c_1Z_\mu+d_1A_\mu,\crn a^2_\mu
&=&b_2Z^\prime_\mu+c_2Z_\mu+d_2A_\mu,\crn a^3_\mu
&=&b_3Z^\prime_\mu+c_3Z_\mu+d_3A_\mu,\label{a123}\eea with\bea
b_1(X_\phi)&=&b_2(X_\phi)=i\fr{g}{6}\fr{3+(6X_\phi-1)t^2_W}{\sqrt{3-t_W^2}},\crn
b_3(X_\phi )&=&i\fr{g}{3}\fr{(3X_\phi
+1)t^2_W-3}{\sqrt{3-t^2_W}},\crn
c_1(X_\phi)&=&i\fr{g}{2c_W}\left[1-2s^2_W\left(\fr{1}{3}+X_\phi\right)\right],\crn
c_2(X_\phi)&=&-i\fr{g}{2c_W}\left[1-2s^2_W\left(\fr{2}{3}-X_\phi\right)\right],\crn
c_3(X_\phi)&=&-ig\fr{s^2_W}{c_W}\left(X_\phi+\fr{1}{3}\right),\crn
d_1(X_\phi)&=&igs_W\left(X_\phi+\fr{1}{3}\right),\crn
d_2(X_\phi)&=&igs_W\left(X_\phi-\fr{2}{3}\right),\crn
d_3(X_\phi)&=&igs_W\left(X_\phi+\fr{1}{3}\right), \eea where
$X_\phi$ is the $X$-charge of the field $\phi$. To find the triple
couplings of gauge bosons with Higgs we expand kinetic terms
(\ref{dnh}):
\begin{widetext}
\bea (D_\mu\phi )^+(D^\mu\phi )&=&[\partial_\mu\phi^++\phi^+
\mbox{diag}(a^1,a^2,a^3)^+_\mu +\phi^+M_\mu^+] [\partial^\mu\phi+
\mbox{diag}(a^1,a^2,a^3)^\mu\phi +M^\mu\phi]\crn
&=&[\partial_\mu\phi^+ \mbox{diag}(a^1,a^2,a^3)^\mu\phi+
H.c.]+\phi^+ \mbox{diag}(a^1,a^2,a^3)^+_\mu
\mbox{diag}(a^1,a^2,a^3)^\mu\phi\crn &+&[\phi^+
\mbox{diag}(a^1,a^2,a^3)^+_\mu M^\mu\phi+H.c.]+\cdots\label{dnh1}
 \eea
 \end{widetext}
 Expression  (\ref{dnh1}) gives the triple couplings among
$Z^\prime$ with  two Higgs bosons and  couplings  among $Z^\prime$
with one Higgs and one gauge  bosons. The couplings among
$Z^\prime$ with two Higgs bosons are given by \bea
a^{i\mu}\partial_\mu\phi^{i*}\phi^{i}+
H.c=b_iZ^{\prime\mu}\partial_\mu\phi^{i*}\phi^{i}+H.c.\label{ttzhh}
\eea where $\phi^T=<\phi^1, \phi^2, \phi^3>$,  the  sum is taken
over $i=1,2,3$ and over $\phi =\eta ,\rho ,\chi$.

The couplings among  $Z^\prime$ with  one  Higgs and one gauge
bosons are given by \bea a^{i*}_\mu
a^{i\mu}\phi_i^*\phi_i+(\phi_i^*a^{i*}_\mu
M^\mu_{ij}\phi^j+H.c.),\label{ttzhg1} \eea where the  sum is taken
over $i=1,2,3$. We see that the necessary couplings are
proportional to VEVs.  Keeping this in mind, from (\ref{ttzhg1}),
we get the following couplings \ben \item For
 $ \phi  \equiv \eta$, with
$
  X_\eta
=-\fr{1}{3}$ and we get \bea a^{1*}_\mu
a^{1\mu}\eta^{0*}\eta^{0}&+&v[(a^{1*}_\mu
M_{1j}^\mu\eta^j+\eta_i^*a^{i*}_\mu M^\mu_{i1}
)+H.c.],\label{ttzhg2} \eea
where $i,j=2,3$.

\item
For  $ \phi \equiv \rho$, with $
  X_\rho
= \fr{2}{3}$ and we get \bea a^{2*}_\mu
a^{2\mu}\rho^{0*}\rho^{0}&+&u[(a^{2*}_\mu
M_{2j}^\mu\rho^j+\rho_i^*a^{i*}_\mu M^\mu_{i2}
)+H.c.],\label{ttzhg3} \eea where $i,j=1,3$. \item For $ \phi
\equiv \chi$, with $
 X_\chi = -\fr{1}{3}$ and one gets
 \bea a^{3*}_\mu
a^{3\mu}\chi^{\prime 0*}\chi^{\prime 0}&+&w[(a^{3*}_\mu
M^\mu_{3j}\chi^j+\chi_i^*a_\mu^{i*}M^\mu_{i3})+H.c.],\crn
\label{ttzhg4}
\eea
where $i,j=1,2$. \een

To get couplings among  $Z^\prime$ with  Higgs bosons we have to
determine the physical Higgs bosons, i.e., we have to consider the
Higgs potential \bea V_{Higgs}&=&\mu_1^2\eta^+\eta
+\mu_2^2\rho^+\rho +\mu_3^2\chi^+\chi
+\lambda_1(\eta^+\eta)^2\crn&+&\lambda_2(\rho^+\rho)^2
+\lambda_3(\chi^+\chi)^2+(\eta^+\eta)[\lambda_4(\rho^+ \rho)\crn
&+&\lambda_5(\chi^+\chi)]
+\lambda_6(\rho^+\rho)(\chi^+\chi)+\lambda_7(\rho^+\eta)(\eta^+\rho)\crn
&+& \lambda_8(\chi^+\eta)(\eta^+\chi)
+\lambda_9(\rho^+\chi)(\chi^+\rho)\crn
&+&[f_1\epsilon^{ijk}\eta_i\rho_j\chi_k+H.c.]. \label{tnh} \eea In
the above potential we have neglected lepton-number violating
interactions since these terms must be very small in comparison
with the above terms, and this does not affect our further results
much.

With VEVs given in  (\ref{tbck}),  we expand the Higgs fields as
follows\footnote{In numerical calculation, one should change to
normalized  fields such as:$\sqrt{2} \eta^0 = v + a_\eta+ib_\eta$}
\bea \eta^0=v+a_\eta+ib_\eta ,\crn \rho^0=u+a_\rho+ib_\rho ,\crn
{\chi^\prime}^0=w+a_\chi+ib_\chi .\label{kth} \eea

Substituting  (\ref{kth}) into the potential  (\ref{tnh}) we see
the mixing between Higgs fields.\footnote{ The mixing happens
between  only the fields having the same  electric charges} After
diagonalization, with the approximation: $|f_1|\sim w$, $w\gg
u,v$\cite{lka1}, we get the following physical fields \bea \left(
\begin{array}{c}
               a_\eta\\ a_\rho
\end{array}  \right)& \approx &\fr{1}{\sqrt{v^2+u^2}}
\left(\begin{array}{cc}v & u\\ u & -v\end{array}\right) \left(
\begin{array}{c}
               H_1^0\\ H_2^0
\end{array}  \right), \crn
a_\chi &\approx &H_3^0,\crn \left( \begin{array}{c}
               b_\eta\\ b_\rho
\end{array}  \right)& \approx &-\fr{1}{\sqrt{v^2+u^2}}\left(
\begin{array}{cc}v & u\\ u & v\end{array}\right)\left( \begin{array}{c}
               G_1^0\\ G_2^0
\end{array}  \right), \crn
b_\chi &\approx &G_3^0.\label{tlh1} \eea with corresponding
masses: \bea m_{H^0_1}^2&\approx
&4\fr{\lambda_2u^4-\lambda_1v^4}{v^2-u^2}\approx
4\lambda_1(u^2+v^2),\crn m_{H^0_2}^2&\approx
&\fr{v^2+u^2}{2vu}w^2,\crn m_{H^0_3}^2&\approx
&-4\lambda_3w^2.\label{klh1} \eea The sectors of  single positive
charged and neutral Higgs fields  give the following states \bea
\left(
\begin{array}{c}
               \eta^+\\ \rho^+
\end{array}  \right)& = &\fr{1}{\sqrt{v^2+u^2}}\left(\begin{array}{cc}
-v & u\\ u & v\end{array}\right)\left( \begin{array}{c}
               G^+_1\\ H_1^+
\end{array}  \right), \crn
\left( \begin{array}{c}
               {\rho^\prime}^+\\ \chi^+
\end{array}  \right)& = &\fr{1}{\sqrt{u^2+w^2}}\left(\begin{array}{cc}
-u & w\\ w & u\end{array}\right)\left( \begin{array}{c}
               G^+_2\\ H_2^+
\end{array}  \right), \crn
\left( \begin{array}{c}
               {\eta^\prime}^0\\ \chi^{0*}
\end{array}  \right)& = &\fr{1}{\sqrt{v^2+w^2}}\left(\begin{array}{cc}
-v & w\\ w & v\end{array}\right)\left( \begin{array}{c}
               G^0_4\\ H_4^0
\end{array}  \right),\label{tlh2}
\eea  with corresponding masses \bea
m_{H_1^+}^2=\fr{v^2+u^2}{2vu}(f_1w-2\lambda_7vu),\crn
m_{H_2^+}^2=\fr{u^2+w^2}{2uw}(f_1v-2\lambda_9uw),\crn
m_{H_4^0}^2=\fr{v^2+w^2}{2vw}(f_1u-2\lambda_8vw).\label{klh2} \eea
Looking at mass spectrum (\ref{klh1}) and (\ref{klh2}) we see that
if  $|f_1|\sim w$, the spectrum separates into two individual
blocks: one very light Higgs $H_1^0$. This field, namely, is the
unique Higgs boson $h$ in the SM and five other Higgs bosons
having  approximately the same masses and proportional to $w$.
Thus, the suggestion $|f_1|\sim w$  is very natural, then we only
need to introduce one new mass scale
 $w$ to extend group
 $SU(2)_L\otimes U(1)_Y\rightarrow
SU(3)_L\otimes U(1)_X$. In the future numerical study, we will
take mass of the $H_1^0\equiv h$ to be about 150 GeV and masses of
all remain Higgs to be in the range of 600 GeV. Note that the
Goldstone bosons, which are $G_{1,2}^0$, $G_3^0$, $G_1^+$, $G_2^+$
and $G_4^0$, correspond to gauge bosons  $Z$, $Z^\prime$, $W^+$,
$Y^+$ and $X$, respectively. The massive states $H_1^0$, $H_2^0$,
$H_3^0$, $H_1^+$, $H_2^+$ and $H_4^0$ are physical ones. From
couplings (\ref{ttzhh}) and mixing (\ref{tlh1}) and (\ref{tlh2}),
in the approximation $w \gg u\approx v$ we have coupling constants
given in Table \ref{zphh}\footnote[1]{The states of Higgs bosons
are hereafter normalized.}.
\begin{table}[h]
\caption{
     The couplings of $Z'$ with two Higgs bosons}
\begin{center}
\begin{tabular}{|c|c|}  \hline
Vertex & Coupling constant ($g_{HH^*}$) \\
\hline $Z^\prime_\mu(\partial^\mu H_1^+H_1^{-}-
\partial^\mu H_1^-H_1^+)$ & $-i\fr{g}{2}\fr{t^2_W}{\sqrt{3-t^2_W}}$\\
\hline $Z^\prime_\mu(\partial^\mu H_2^-H_2^{+}-
\partial^\mu H_2^+H_2^-)$ & $-ig\fr{1-t^2_W}{\sqrt{3-t^2_W}}$\\
\hline $Z^\prime_\mu(\partial^\mu H_4^{0*}H_4^{0}-
\partial^\mu H_4^{0}H_4^{0*})$ & $-ig\fr{1}{\sqrt{3-t^2_W}}$\\
\hline
\end{tabular}\label{zphh}
\end{center}
\end{table}

Similarly, from  interactions (\ref{ttzhg2}) -(\ref{ttzhg4})
 and mixing (\ref{tlh1}) and
(\ref{tlh2}) we get
 the coupling constants of $Z'$ to one Higgs and another
gauge bosons which are given in Table \ref{zphg}.

\begin{table*}
\caption{
     Triple coupling constants of $Z'$, gauge  and  Higgs bosons }
\begin{center}
\begin{tabular}{|c|c|c|c|}  \hline
Vertex & Coupling constant [$(g_{HG})g_{\alpha\beta}$]&Vertex &
Coupling constant [$(g_{HG})g_{\alpha\beta}$]\\
\hline $Z^\prime_\mu Z^\mu H_1^{0}$
&$ug^2\fr{(1-t^2_W)}{\sqrt{3-t^2_W}}$&$
Z^\prime_\mu {Z^\prime}^\mu H_1^{0}$  &$ug^2\fr{1+t^4_W}{3-t^2_W}$\\
\hline $Z^\prime_\mu Z^\mu H_2^{0}$  &$0$&$Z^\prime_\mu
{Z^\prime}^\mu H_2^{0}$  &$
-ug^2\fr{2t^2_W}{3-t^2_W}$\\
\hline $Z^\prime_\mu Z^\mu H_3^{0}$  &$0$&$Z^\prime_\mu
{Z^\prime}^\mu
 H_3^{0}$  &$wg^2\fr{2\sqrt{2}}{3-t^2_W}$\\
\hline $Z^\prime_\mu X^\mu H_4^{0}$
&$vg^2\fr{\sqrt{3-t^2_W}}{\sqrt{2}}$&$
Z^\prime_\mu Y^{-\mu} H_2^{+}$  &$ug^2\fr{\sqrt{3-t^2_W}}{\sqrt{2}}$\\
\hline $Z^\prime_\mu {X^*}^\mu H_4^{0*}$
&$vg^2\fr{\sqrt{3-t^2_W}}{\sqrt{2}}$&$
Z^\prime_\mu Y^{+\mu} H_2^{-}$  &$ug^2\fr{\sqrt{3-t^2_W}}{\sqrt{2}}$\\
\hline
\end{tabular}\label{zphg}
\end{center}
\end{table*}

From Table \ref{zphg}, we see that the coupling constants are
always proportional to VEV and $g^2$. In addition, the decay
channels  $Z^\prime\to Zh$ at high energy may give contribution to
production of the SM  Higgs boson $h$. This background
contribution is not small in comparison with the discovery
threshold at LHC (this will be shown in our conclusions) and it is
necessary to be determined in the process of searching the Higgs
boson in the near future. The channel ${Z^\prime}^*\to Z^\prime h$
gives the possibility to the process $pp\to Z^\prime h$. The cross
section for this process, however, must be phenomenologically
quite small. We will consider both processes in our numerical
evaluation.

To proceed further, let us give masses of the gauge bosons after
symmetry breaking \cite{hnl}
 \bea M_W^2&=&\fr{1}{2}g^2(u^2+v^2),\
M_Z^2=\fr{g^2}{2c^2_W}(u^2+v^2),\crn
M_X^2&=&\fr{1}{2}g^2(w^2+v^2),\ M_Y^2=\fr{1}{2}g^2(w^2+u^2),\crn
M_{Z^\prime}^2&=&\fr{g^2}{2(3-4s^2_W)}\left[4w^2
+\fr{u^2}{c^2_W}+\fr{v^2(1-2s^2_W)^2}{c^2_W}\right].\label{mgb}
\eea
For  simplicity we suggest that
 \bea
w\gg u\approx v \approx \fr{M_W}{g}. \eea Then, from (\ref{mgb})
we get the following relation \bea M_{Y}\approx M_X\approx
\fr{\sqrt{3-4s^2_W}}{2}M_{Z^\prime}\approx
0.72M_{Z^\prime}.\label{mxyz} \eea  Hence, by (\ref{mxyz}), the
processes such as  $Z^\prime\to XX^*, Y^-Y^+$ do not give
contribution to decay width of $Z^\prime$. In addition, from the
rare kaon decay, the following constraint is given \cite{vanlong}
 \bea 2.3 \ \mbox{TeV} < M_{Z^\prime} < 4.3\ \mbox{TeV}. \eea
In our numerical study, we will release the lower limit to be 1
TeV. From (\ref{mxyz}) we get \bea 1.6 \ \mbox{TeV} < M_{X}\approx
M_Y < 3.1\ \mbox{TeV}. \eea

\section{Decay width of $Z^\prime$}

In the processes mediated by  the $Z$ and $Z'$  neutral gauge
bosons, the decay widths of these particles play  very important
role. Hence, we firstly discuss of decay modes of these particles.
 Until  now, the total decay width of  $Z^\prime$  into
particles in the 3-3-1 model with RH neutrinos are not exactly and
completely performed. In \cite{cmt1,ptt}, the two-body $Z^\prime$
decay in the 3-3-1 model with exotic leptons is considered at tree
level but not complete. \emph{The partial decay of $Z^\prime$ into
$H^+Y^-$ is absent there}. Since the value $\Gamma_{Z^\prime}$ has
very important in hadron collisions
 $ p p$ with intermediate field  $Z^\prime$, therefore
we try to give the exact value for
  $\Gamma_{Z^\prime}$ at the tree level.
Generally we have \bea \Gamma_{Z^\prime}&=&\Gamma (Z^\prime\to\nu
\bar{\nu})+\Gamma (Z^\prime\to l\bar{l})+\Gamma (Z^\prime\to q\bar
q)\crn &+&\Gamma (Z^\prime\to Q\bar Q)+\Gamma (Z^\prime\to
GG^*)+\Gamma (Z^\prime\to HH^*)\crn &+&\Gamma (Z^\prime\to HG)
\eea where $q, Q, G, H$ stand for the usual light quarks of the
SM, exotic quarks,  gauge  and  Higgs bosons, respectively.

By the couplings  (\ref{ncva}) the decay of $Z'$ into leptons has
a general form \cite{ptt}
\begin{widetext}
 \bea \Gamma
(Z^\prime\to
f\bar{f})=m_{Z^\prime}\fr{N_C}{12\pi}\fr{g^2}{c^2_W}\sqrt{1
-\fr{4m^2_f}{m_{Z^\prime}^2}}\left[|g_{2V}(f)|^2\left(1+
\fr{2m^2_f}{m_{Z^\prime}^2}\right)+|g_{2A}(f)|^2\left(1-
\fr{4m^2_f}{m_{Z^\prime}^2}\right)\right],\label{kzff} \eea
\end{widetext}where
$N_C$  is the color number of $f$. In this work, masses of leptons
are neglected since they are very small in comparison with the
$Z'$ mass. Masses of the exotic quarks are assumed to be equal and
are in the range of 600 GeV.

Because of (\ref{mxyz}), it follows that $\Gamma (Z^\prime\to G
G^*)=0$. The two Higgs boson decay gets a general form
 \bea \Gamma (Z^\prime\to
HH^*)=m_{Z^\prime}\fr{|g_{HH^*}|^2}{48\pi}
\left(1-4\fr{m_H^2}{m_{Z^\prime}^2}\right)^{3/2}.\label{kzhh} \eea
 Note that (\ref{kzhh}) is valid only for the case
where two Higgs bosons have the equal masses. Finally, for the
decay channels with one Higgs boson and one gauge particle, we
have \bea \Gamma (Z^\prime\to
HG)=\fr{|g_{HG}|^2}{24\pi}\fr{\sqrt{E_G^2-m_G^2}}{m_{Z^\prime}^2}
\left(2+\fr{E_G^2}{m_{G}^2}\right),\label{kzhg} \eea where $E_G$
is energy of the gauge boson at the final state given by $E_G =
(m_{Z^\prime}^2+m_G^2-m_H^2)/2m_{Z^\prime}$. The total decay width
of $Z'$ in the 3-3-1 model with right-handed neutrinos is plotted
in Fig. \ref{gzpflt}.
\begin{figure*}[htbp]
\begin{center}
\includegraphics[width=12cm,height=8cm]{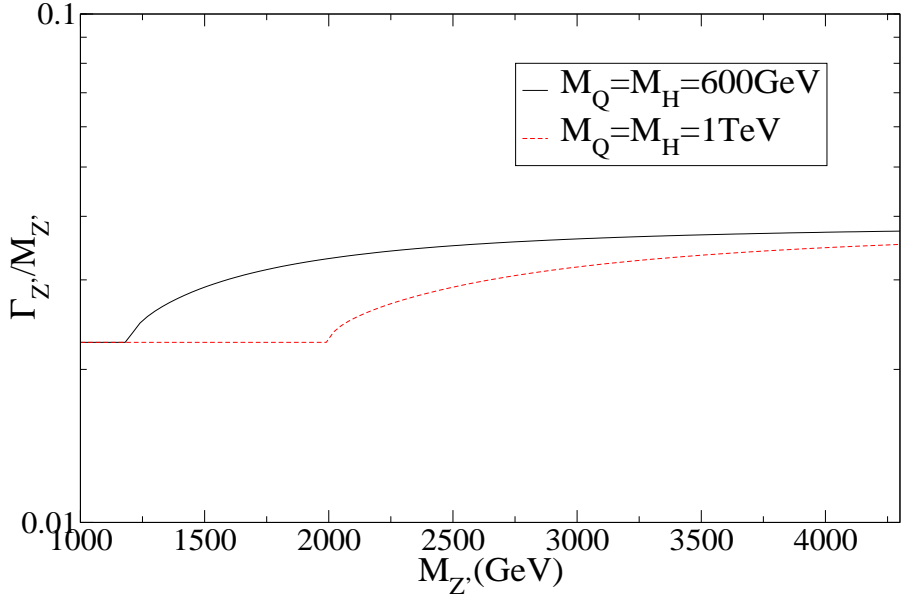}
\caption{\label{gzpflt}{\em $\Gamma_{Z^\prime}/M_{Z^\prime}$ as a
function of $M_{Z^\prime}$ in 3-3-1 model with RH neutrinos.}}
\end{center}
\end{figure*}

The branching ratios as function of the $Z'$ mass are plotted in
Fig. \ref{ratio}.
\begin{figure}[htbp]
\begin{center}
\includegraphics[width=8cm,height=6cm]{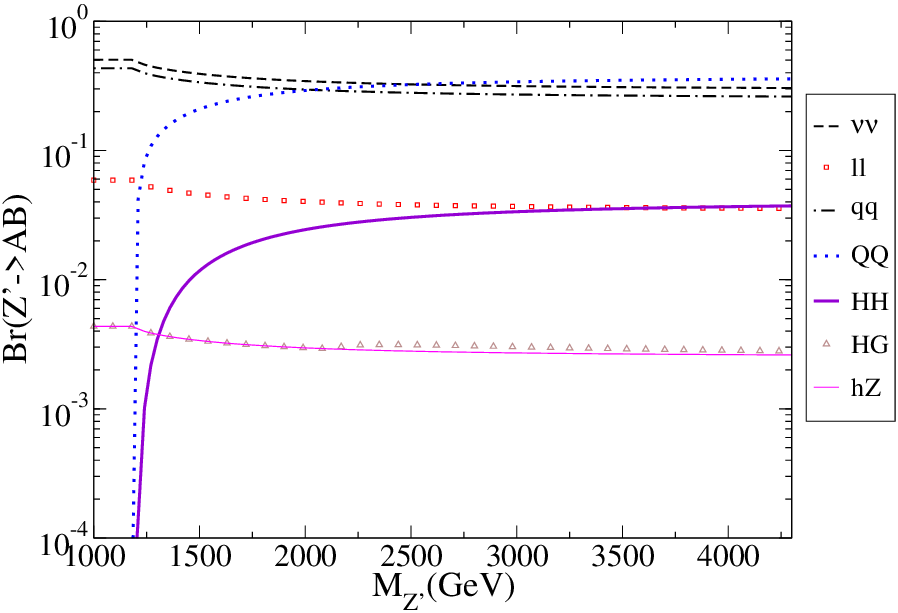}
\caption{\label{ratio}{\em Branching ratios of $Z^\prime$ in 3-3-1
model with RH neutrinos in the scenario $M_Q=M_H=600$ GeV,
$M_{h}=150$ GeV.}}
\end{center}
\end{figure}
\section{Production of $Z'$ and SM-like Higgs  boson at LHC}
In this section we consider production of extra neutral gauge
boson $Z'$ and its contribution to the Higgsstrahlung in the 3-3-1
model with RH neutrinos at hadron colliders, namely, the single
$Z'$ production $ p p \to Z^\prime$, $ p p \to Zh$ and  $ p p \to
Z^\prime h$. We first turn on the single $Z'$ production.
\subsubsection{Single $Z'$ production at LHC}
In contract with the lepton colliders  like $e^+ e^-$, it is
possible the single production of the $Z$ and $Z'$ at hadron
colliders.  Let us consider the available fusion subprocess
   \bea q + \bar{q} \to Z^\prime
\eea which is dominantly by $q \bar{q}$ annihilation, and the
resulting cross-section in given by
 \bea \hat\sigma(q\bar q\to Z^\prime
)=\fr{1}{3}\fr{\pi g^2}{c^2_W}[(g^q_{2V})^2+(g^q_{2A})^2]\delta
(\hat s-M_{Z^\prime}^2) \eea where
 \bea \hat s&=&x_1x_2s.\label{xe}\eea
This subprocess in the minimal 3-3-1 model  was considered in
\cite{fr1}.
 Because of (\ref{xe}) we have
\bea \delta (\hat s-M_{Z^\prime}^2)&=&\fr{1}{sx_2}\delta
\left(x_1-\fr{M_{Z^\prime}^2}{sx_2}\right). \eea The total
cross-section of the scattering process $p p\to Z^\prime X$ is
given by
\begin{widetext}
 \bea \sigma( p p \to Z^\prime
)&=&2\sum_{i=1}^5\int_0^1dx_1\int_0^1dx_2f(i,x_1,Q)f(-i,x_2,Q)\hat\sigma(q_i\bar
q_i\to Z^\prime )\crn &=&\fr{2\pi
g^2}{3c^2_W}\fr{1}{s}\sum_{i=1}^5[(g_{2V}^{qi})^2+(g_{2A}^{qi})^2]
\int_{M_{Z^\prime}^2/s}^1f(i,x_1,M_{Z^\prime})f\left(-i,
\fr{M_{Z^\prime}^2}{sx_1},M_{Z^\prime}\right)\fr{dx_1}{x_1}.\label{totalcr}
\eea \end{widetext} Using  CTEQ6M (2004) \cite{cteq6} where
$t$-quark is not included, we obtain cross-section for the above
process which is plotted in Fig. \ref{sing}.
\begin{figure}[t]
\begin{center}
\includegraphics[width=8cm,height=6cm]{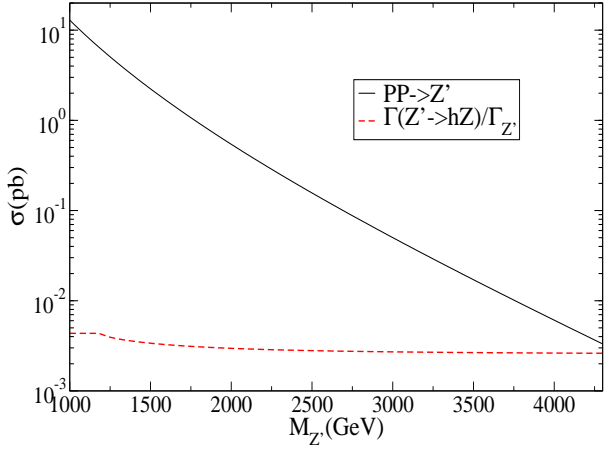}
\caption{\label{sing}{\em Single $Z'$ production in 3-3-1 model
with RH neutrinos at LHC with $\sqrt{s}=14\ \mbox{TeV}$. The
dotted line indicates the $Z^\prime$ branching ratio into $hZ$.}}
\end{center}
\end{figure}

 \subsection{Production of the SM-like Higgs  boson at LHC}
Next, we consider production of the SM  Higgs boson at hadron
colliders due to the $Z'$ in the intermediate state, namely
 \bea
q(p_1)\bar{q}(p_2)\to Z, Z^\prime \to Z(k_1)h(k_2).\label{ppzh1}
\eea The differential cross-section is given by
\begin{widetext}
\bea
\fr{d\hat{\sigma}}{d\cos\theta}&=&\fr{1}{N_C}\fr{\hat{\beta}_hg^2}{64\pi
c_W^2}\fr{1}{\hat{s}}
\left\{\fr{(g_{ZZh})^2}{(\hat{s}-M_{Z}^2)^2+(M_{Z}\Gamma_{Z})^2}[\hat{s}
+\fr{(M_Z^2-\hat{t})(M_Z^2-\hat u)}{M_Z^2}\right ][
(g_{1V}^{qi})^2+(g_{1A}^{qi})^2]\nonumber\\
 &+& \left.\fr{(g_{Z'Zh})^2}{(\hat{s}-M_{Z'}^2)^2+(M_{Z'}\Gamma_{Z'})^2}
            [\hat{s}+\fr{(M_Z^2-\hat{t})(M_Z^2-\hat u)}{M_Z^2}\right]
[(g_{2V}^{qi})^2+(g_{2A}^{qi})^2]\crn &+& \left.
2Re[\fr{(g_{ZZh})(g_{Z'Zh})^*}{(\hat{s}-M_Z^2+iM_Z\Gamma_Z)(
\hat{s}-M_{Z'}^2-iM_{Z'}
\Gamma_{Z'})}][\hat{s}+\fr{(M_Z^2-\hat{t})(M_Z^2-\hat
u)}{M_Z^2}\right]\crn &\times& \left.
[g_{2V}^{qi}g_{1V}^{qi}+g_{2A}^{qi}g_{1A}^{qi}]\right\}, \eea
where $N_C=3$ is the number of colors, $1/N_C=3\times 1/3\times
1/3$ is the averaging over colors, \bea\hat{s}&=&x_1x_2s,\crn
\hat{\beta}_h&=&\left(1+\fr{M_h^2-M_Z^2}{\hat
s}\right)\sqrt{1-\fr{4\hat sM_h^2}{(\hat{s}-M_Z^2+M_h^2)^2}},\crn
\hat{t}&=&M_Z^2-\fr{\hat s+M_Z^2-M_h^2}{2}
\left[1-\cos\theta\sqrt{1-\fr{4\hat
sM_Z^2}{(\hat{s}+M_Z^2-M_h^2)^2}}\right],\nn \eea
\bea\hat{u}&=&M_h^2-\fr{\hat s-M_Z^2+M_h^2}{2}
\left[1+\cos\theta\sqrt{1-\fr{4\hat
sM_h^2}{(\hat{s}-M_Z^2+M_h^2)^2}}\right],\eea
\end{widetext}
\bea g_{ZZh}&=&\fr{g}{c^2_W}M_W,\crn g_{Z^\prime
Zh}&=&gM_W\fr{1-t^2_W}{\sqrt{3-t^2_W}}. \eea

 After some
manipulation, this formula is similar to Eq. (10) in \cite{cmt1}.
The total cross-section for the above process has a form
\begin{widetext}
  \bea
\sigma(p p\to hZ )&=&2\sum_{i=1}^5\int_{-1}^{1}d(\cos\theta)
\int_0^1dx_1\int_0^1dx_2f(i,x_1,Q)f(-i,x_2,Q)\fr{d\hat\sigma(q_i\bar
q_i\to hZ )}{d\cos\theta}.\crn \eea
\end{widetext}
The total cross-section of the process as a function of the $Z'$
mass is plotted in Fig. \ref{zhp} \vs
\begin{figure}[b]
\begin{center}
\includegraphics[width=8cm,height=6cm]{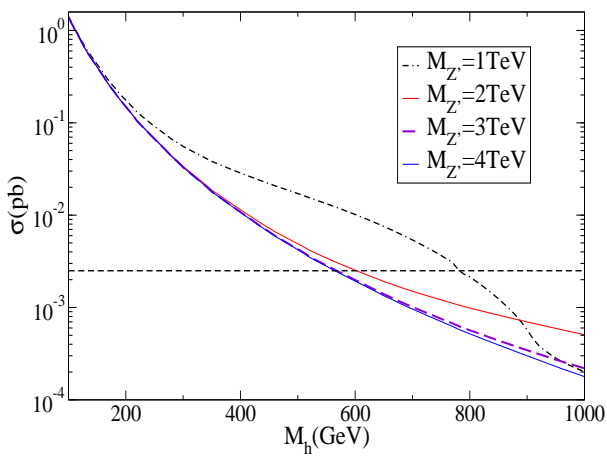}
\caption{\label{zhp}{\em Total cross section for the process $p
p\to hZ$ as a function of $M_{h}$ for various values of
$M_{Z^\prime}$ at LHC  with $\sqrt{s}=14 \mbox{ TeV}$ in 3-3-1
model with RH neutrinos. The horizontal line indicates the cross
section required for discovery: 2.5 fb.}}
\end{center}
\end{figure}
The discovery limit is taken from  Ref.\cite{london}.

It is instructive to separate out the contribution of the on-shell
$Z^\prime$. And from these results we can estimate the correction
of $Z'$ in the production of SM  Higgs at LHC (see Fig.
\ref{rzhp}).\vs

\begin{figure}[t]
\begin{center}
\includegraphics[width=8cm,height=6cm]{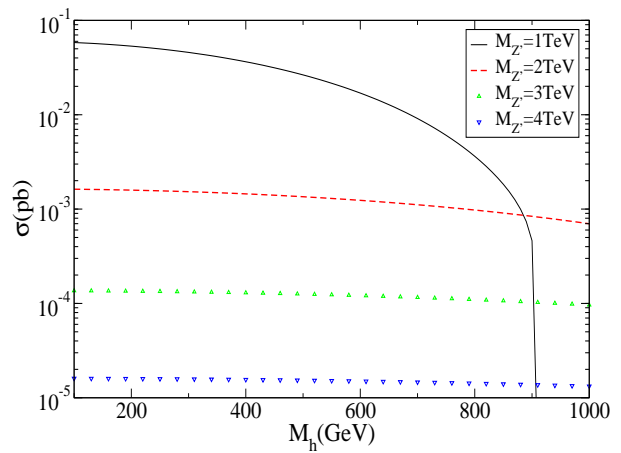}
\caption{\label{rzhp}{\em The real $Z^\prime$ contribution to $ h
Z$ production at LHC as a function of $M_h$ for various values of
$M_{Z^\prime}$.}}
\end{center}
\end{figure}

We also consider the production of $hZ^\prime$ pair as promised at
LHC. Of course, this cross section should be very small. It is
impossible to observe this production at LHC (see Fig.
\ref{z2hp}).

\begin{figure}[b]
\begin{center}
\includegraphics[width=8cm,height=6cm]{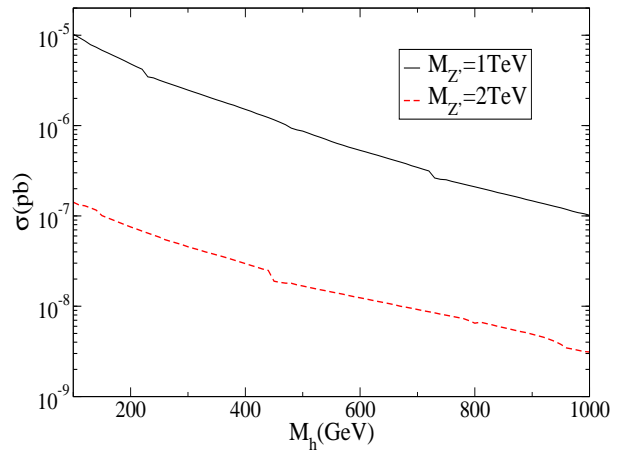}
\caption{\label{z2hp}{\em Total cross section for the process $p
p\to hZ^\prime$ as a function of $M_{h}$ for various values of
$M_{Z^\prime}$ at LHC  with $\sqrt{s}=14\ \mbox{ TeV}$ in 3-3-1
model with RH neutrinos.}}
\end{center}
\end{figure}
\section{Conclusions}

We have investigated the production of  neutral Higgs boson at
hadron colliders. Here are our results: \ben
\item The  $Z'$ decay modes including contribution from Higgs bosons are
obtained. The branching ratios for the decay modes $Z^\prime \to
\nu\bar{\nu};\ l\bar{l};\ $ $ q\bar{q};\ Q\bar{Q}$\ $
(D_1\bar{D_1}$, $D_2\bar{D_2}$, $T\bar{T})$; $ HH^*$$(H_1^+H_1^-$,
$H_2^+H_2^-,H_4^0H_4^{0*})$ and $HG
$($H_1^0Z$,$H_4^0X$,$H_4^{0*}X^*$,$H_2^+Y^-$,$H_2^-Y^+$) are
displayed in Fig. \ref{ratio} as a function of $M_{Z^\prime}$ in
the scenario $M_Q=M_H=600$ GeV, $M_{h}=150$ GeV. In the region
$M_{Z^\prime}\in [1,1.2\ \mbox{TeV}]$  the partial decays of
$Z^\prime$ into $Q\bar{Q}$ and $HH^*$ are kinetically not allowed
and $Br(Z^{\prime}\to \nu\bar{\nu})\approx 0.5$, $Br(Z^{\prime}\to
q\bar{q})\approx 0.43$ are dominant. $Br(Z^\prime\to l\bar{l})$ is
of the order $10^{-2}$ and the mode $Z^\prime\to HG$ is marginal,
with a branching ratio of the order $10^{-3}$. When the mass of
$Z^\prime$ is higher than 1.2 TeV the decay modes of $Z^\prime$
into $Q\bar{Q}$ and $HH^*$ start and give considerable
contributions. The branching ratio $Br(Z^\prime\to
Q\bar{Q})\approx 0.36$, a bit higher than $Br(Z^\prime\to
\nu\bar{\nu})$ which slopes slightly down when $M_{Z^\prime}$
increases. The partial width of $Z^\prime$ into heavy Higgs cannot
be neglected and $Br(Z^\prime\to HH^*)\approx Br(Z^\prime\to
l\bar{l})$ which is nearly unchanged.

The two-body decay width of $Z^\prime$ is calculated and the value
of $\Gamma_{Z^\prime}/M_{Z^\prime}$ is shown in Fig. \ref{gzpflt}
as a function of $M_{Z^\prime}$ in two scenarios $M_Q=M_H=600$
GeV, $M_Q=M_H=1$ TeV and $M_{h}=150$ GeV in both cases.
$\Gamma_{Z^\prime}$ is typically about $2\div 4\%$ $M_{Z^\prime}$.
We emphasize here that if the constraint (\ref{mxyz}) is released
as in \cite{cmt1,ptt} and if we take $M_X=M_Y=600$ GeV then the
decay modes $Z^\prime\to XX^*,Y^-Y^+$ are allowed. As a
consequence, $\Gamma_{Z^\prime}$ will become very large to the
values 413.74 GeV and 2.6 TeV, for the cases $M_{Z^\prime}$ are 2
TeV and 3 TeV, respectively.
\item Cross section for the fusion subprocess (single $Z'$
production) is presented in Fig. \ref{sing}.
\item Production the the SM-like Higgs  boson is given in Fig. \ref{zhp}.
The design luminosity at LHC is 10 $fb^{-1}/yr$. We require 25
events for discovery as in \cite{london}. This corresponds to a
cross section of 2.5 fb. If the mass of $Z^\prime$ is not higher
than 1 TeV then the process $pp\to hZ$ is observable at LHC in the
case $M_{h}< 780$ GeV. If $M_{Z^\prime}$ is from $2\div 4$ TeV
then the process is observable, for $M_{h}< 600$ GeV. At the
present constraint: $M_{h}=114_{-40}^{+56}$ \cite{pdg}, we see
from Fig. 4 and  5 that the SM-like Higgs is mainly produced via
$Z$ exchange. The contribution of the real $Z'$ is about
$10^{-3}\div 10^{-1}$ pb if $1\ \mbox{TeV}\leq M_{Z^\prime}\leq 2\
\mbox{TeV}$  and it can be neglected in the case $M_{Z^\prime}$ is
above that range.
\item The total cross section for the process $p p\to hZ^{\prime}$
is also displayed in Fig. \ref{z2hp}. The cross section is about
0.01 fb, for $M_{Z^\prime}=1$ TeV. This process is unobservable at
LHC. \een

To finish, we hope that our results will make Higgs structure
more clear when the LHC  is in operation.\\

\begin{acknowledgments}

\hs One of the authors, Le Duc Ninh, would like to express his
gratefulness to Patrick Aurenche for teaching him the techniques
of hadron collisions and for many fruitful discussions. This work
is supported in part by National Council for Natural Sciences of
Vietnam contract No: KT - 41064.
\end{acknowledgments}

\end{document}